\documentclass[12pt]{article}
\newcommand{\be}{\begin{equation}}
\newcommand{\ee}{\end{equation}}

\usepackage{epsfig,amsmath,amssymb,graphics,graphicx} 
\newcommand{\bp}{{\bf Proof.} }
\newcommand{\ep}{$\ \ \ \Box$ \\ }

\begin{document}

\begin{center}
Chapter 1, in {\it Long-range Interaction, Stochasticity and Fractional Dynamics}, \\
A.C.J. Luo, V. Afraimovich, (Eds.), (Springer, HEP, 2010) pages 1-26.
\end{center}

\begin{center}
{\Large \bf  Fractional Zaslavsky and H\'enon Discrete Maps}
\vskip 5 mm
{\large \bf Vasily E. Tarasov}\\
\vskip 3mm
{\it Skobeltsyn Institute of Nuclear Physics, \\
Moscow State University, Moscow 119991, Russia } \\
{E-mail: tarasov@theory.sinp.msu.ru}
\vskip 11 mm

\begin{abstract}
This paper is devoted to the memory of Professor George M. Zaslavsky 
passed away on November 25, 2008.
In the field of discrete maps, George M. Zaslavsky 
introduced a dissipative standard map which is called now the Zaslavsky map. 
G. Zaslavsky initialized many fundamental concepts and 
ideas in the fractional dynamics and kinetics. 
In this paper, 
starting from kicked damped equations with derivatives of non-integer orders 
we derive a fractional generalization of discrete maps.
These fractional maps are generalizations of the Zaslavsky map and the H\'enon map.
The main property of the fractional differential equations and 
the correspondent fractional maps is a long-term memory and dissipation.
The memory is realized by the fact that their present state evolution depends 
on all past states with special forms of weights. 
\end{abstract}

\end{center}

\section{Introduction}

There are a number of distinct areas of mechanics and physics where
the basic problems can be reduced to the study of simple discrete maps. 
Discrete maps have been used for the study of dynamical problems, 
possibly as a substitute of differential equations 
\cite{Zbook1,Zbook3,Chirikov,Schuster,CE}. 
They lead to a much simpler formalism, which is particularly useful 
in computer simulations.
In this chapter, we consider discrete maps 
that can be used to study the evolution described 
by fractional differential equations \cite{SKM,Podlubny,KST}. 

The treatment of nonlinear dynamics in terms of discrete maps 
is a very important step in understanding the qualitative
behavior of continuous systems described by differential equations.
The derivatives of non-integer orders \cite{SKM} are a natural generalization of 
the ordinary differentiation of integer order. 
Note that the continuous limit of discrete systems with power-law
long-range interactions gives differential equations 
with derivatives of non-integer orders 
with respect to coordinates (see for example, \cite{TZ3,JPA2006}).
Fractional differentiation with respect to time is
characterized by long-term memory effects that 
correspond to intrinsic dissipative processes in the physical systems. 
The memory effects to discrete maps mean 
that their present state evolution depends on all past states. 
The discrete maps with memory are considered in the papers
\cite{Ful,Fick1,Giona,Fick2,Gallas,Stan,JPA2008,JPA2009,PLA2009}. 
The interesting question is a connection of fractional equations  
of motion and the discrete maps with memory. 
This derivation is realized for universal and standard maps in \cite{JPA2008,JPA2009}.

It is important to derive discrete maps with memory 
from equations of motion with fractional derivatives.
It was shown \cite{ZSE} that perturbed by a periodic force, 
the nonlinear system with fractional derivative exhibits a new type 
of chaotic motion called the fractional chaotic attractor. 
The fractional discrete maps \cite{JPA2008,JPA2009} can be used to study 
a new type of attractors that are called pseudochaotic \cite{ZSE}. 

In this capter, fractional equations of motion for kicked systems 
with dissipation are considered. 
Correspondent discrete maps are derived. 
The fractional generalizations of the Zaslavsky map and the H\'enon map
are suggested.

In Sec. 2, we give a brief review of fractional derivatives
to fix notation and provide a convenient reference.
In Sec. 3, the fractional generalizations of the Zaslavsky map are suggested.
A brief review of well-known discrete maps 
is considered to fix notations and provide convenient references. 
In Sec. 4, the fractional generalizations of the H\'enon  map are considered.
The differential equations with derivatives of non-integer orders 
with respect to time are used to derive generalizations of the discrete maps.
In Sec. 5, a fractional generalization of 
differential equation in which we use a fractional derivative 
of the order $0 \le \beta < 1$ in the kicked term, i.e. the term 
of a periodic sequence of delta-function type pulses (kicks).
The other generalization is suggested in \cite{JPA2008}.
The discrete map that corresponds to the suggested fractional equation 
of order $0 \le \beta < 1$ is derived. 
This map can be considered as a generalization of universal map
for the case $0 <\beta < 1$.
In Sec. 6, a fractional generalization of 
differential equation for a kicked damped rotator is suggested. 
In this generalization, we use a fractional derivative in 
the kicked damped term, i.e. the term 
of a periodic sequence of delta-function type pulses (kicks).
The other generalization is suggested in \cite{JPA2008}.
The discrete map that corresponds to the suggested fractional 
differential equation is derived. 
Finally, a short conclusion is given in Sec. 7.

\section{Fractional derivatives}

In this section a brief introduction to fractional derivatives are suggested.
Fractional calculus is a theory of integrals and 
derivatives of any arbitrary order.
It has a long history from 1695, when the derivative of 
order $\alpha=1/2$ has been described by Gottfried Leibniz.
The fractional differentiation and fractional integration 
goes back to many mathematicians such as 
Leibniz, Liouville, Grunwald, Letnikov, Riemann, Abel, Riesz, Weyl. 
The integrals and derivatives of non-integer order, 
and the fractional integro-differential equations have found many applications 
in recent studies in theoretical physics, mechanics and applied mathematics.
There exists the remarkably comprehensive 
encyclopedic-type monograph by Samko, Kilbus and Marichev, 
which was published in Russian in 1987 and in English in 1993.
The works devoted substantially to fractional differential equations 
are the book by Miller and Ross (1993), and the book by Podlubny (1999).
In 2006 Kilbas, Srivastava and Trujillo published a very important and 
remarkable book, where one can find a modern encyclopedic, 
detailed and rigorous theory of fractional differential equations.
The first book devoted exclusively 
to the fractional dynamics and application of fractional calculus
to chaos is the book by Zaslavsky published in 2005.

Let us give a brief review of fractional derivatives
to fix notation and provide a convenient reference.

\subsection{Fractional Riemann-Liouville derivatives}

Let $[a,b]$ be a finite interval on the real axis $R$. 
The fractional Riemann-Liouville derivatives $D^{\alpha}_{a+}$ and $D^{\alpha}_{b-}$ 
of order $\alpha >0$ are defined \cite{KST} by
\[ (D^{\alpha}_{a+}f)(x)=D^n_x (I^{n-\alpha}_{a+})(x)= \]
\[ = \frac{1}{\Gamma(n-\alpha)} D^n_x \int^x_a \frac{f(z)dz}{(x-z)^{\alpha-n+1}} 
\quad  (x>a) , \]
\[ (D^{\alpha}_{b-}f)(x)=(-1)^n D^n_x (I^{n-\alpha}_{b-})(x)= \]
\[ = \frac{(-1)^n}{\Gamma(n-\alpha)} D^n_x \int^b_x \frac{f(z)dz}{(z-x)^{\alpha-n+1}} 
\quad  (x<b) , \]
where $n=[\alpha]+1$ and $[\alpha]$ means the integral part of $\alpha$.
Here $D^n_x$ is the usual  derivative of order $n$. 
In particular, when $\alpha=n \in N$, then
\[ (D^0_{a+}f)(x)=(D^0_{b-}f)(x)=f(x) , \]
\[ (D^n_{a+}f)(x)=D^n_x f(x) , \quad (D^n_{b-}f)(x)=(-1)^n D^n_x f(x) . \]

The fractional Riemann-Lioville differentiation of the power 
functions $(x-a)^{\beta}$ and $(b-x)^{\beta}$ yields power functions 
of the same form
\[ D^{\alpha}_{a+} (x-a)^{\beta}=
\frac{\Gamma(\beta+1)}{\Gamma(\alpha+\beta+1)} (x-a)^{\beta-\alpha} , \]
\[ D^{\alpha}_{b-} (b-x)^{\beta}=
\frac{\Gamma(\beta+1)}{\Gamma(\alpha+\beta+1)} (b-x)^{\beta-\alpha} , \]
where $\beta>-1$ and $\alpha>0$. 
In particular, if $\beta=0$ and $\alpha>0$, 
then the fractional Riemann-Liouville derivatives 
of a constant $C$ are not equal to zero:
\[ D^{\alpha}_{a+} C=\frac{1}{\Gamma(\alpha+1)} (x-a)^{-\alpha} , \]  
\[ D^{\alpha}_{b-} C=\frac{1}{\Gamma(\alpha+1)} (b-x)^{-\alpha} . \]   
On the other hand, for $k=1,2,...,[\alpha]+1$, we have 
\[ D^{\alpha}_{a+} (x-a)^{\alpha-k}=0, \quad D^{\alpha}_{b-} (b-x)^{\alpha-k}=0 . \]

The equality
\[ (D^{\alpha}_{a+}f)(x)=0 \]
is valid if, and only if,
\[ f(x)=\sum^n_{k=1} C_k (x-a)^{\alpha-k} , \]
where $n=[\alpha]+1$ and $C_k$ are real arbitrary constants.
The equation
\[  (D^{\alpha}_{b-}f)(x)=0 \]
is satisfied if, and only if,
\[ f(x)=\sum^n_{k=1} C_k (b-x)^{\alpha-k} , \]
where $n=[\alpha]+1$ and $C_k$ are real arbitrary constants.

\subsection{Fractional Caputo derivative}

The fractional Caputo derivatives $\ ^CD^{\alpha}_{a+}$ and $\ ^CD^{\alpha}_{b-}$
are defined for functions for which the Riemann-Liouville derivatives exists.
Let $\alpha>0$ and let $n$ be given by
$n=[\alpha]+1$ for $\alpha \not \in N$, 
and $n=\alpha$ for $\alpha \in N$.
If $\alpha \not \in N$, then the fractional Caputo derivatives
is defined by the equations
\[ (\ ^CD^{\alpha}_{a+}f)(x) = (I^{n-\alpha}_{a+} D^n f)(x)=
 \frac{1}{\Gamma(n-\alpha)} \int^x_a dz \frac{D^n_zf(z)}{(x-z)^{\alpha-n+1}} , \]
\[ (\ ^CD^{\alpha}_{b-}f)(x) =(-1)^n (I^{n-\alpha}_{b-} D^n f)(x)=
 \frac{(-1)^n}{\Gamma(n-\alpha)} \int^b_x dz \frac{D^n_zf(z)}{(z-x)^{\alpha-n+1}} , \]
where $n=[\alpha]+1$. 
If $\alpha=n \in N$, then
\[ (\ ^CD^{\alpha}_{a+}f)(x) = D^n_x f(x), \quad 
(\ ^CD^{\alpha}_{b-}f)(x) =(-1)^n D^n_x f(x) . \]
If $\alpha \not \in N$ and $n=[\alpha]+1$, 
then fractional Caputo derivatives coincide
with the fractional Riemann-Liouville derivatives in the following cases:
\[ (\ ^CD^{\alpha}_{a+}f)(x) = (D^{\alpha}_{a+}f)(x) , \]
if
\[ f(a)=(D^1_xf)(a)=...=(D^{n-1}f)(a)=0 , \]
and
\[ (\ ^CD^{\alpha}_{b-}f)(x) = (D^{\alpha}_{b-}f)(x) , \]
if
\[ f(b)=(D^1_xf)(b)=...=(D^{n-1}f)(b)=0 . \]

It can be directly verified that the fractional Caputo differentiation of
the power functions $(x-a)^{\beta}$ and $(b-x)^{\beta}$ yields power functions 
of the form
\[ ^CD^{\alpha}_{a+} (x-a)^{\beta}=
\frac{\Gamma(\beta+1)}{\Gamma(\alpha+\beta+1)} (x-a)^{\beta-\alpha} , \]
\[ ^CD^{\alpha}_{b-} (b-x)^{\beta}=
\frac{\Gamma(\beta+1)}{\Gamma(\alpha+\beta+1)} (b-x)^{\beta-\alpha} , \]
where $\beta>-1$ and $\alpha>0$. 
In particular, if $\beta=0$ and $\alpha>0$, then the fractional Caputo derivatives 
of a constant $C$ are equal to zero:
\[  ^CD^{\alpha}_{a+} C=0 , \quad  ^CD^{\alpha}_{b-} C=0. \]   
For $k=0,1,2,...,n-1$, 
\[ ^CD^{\alpha}_{a+} (x-a)^k=0, \quad  D^{\alpha}_{b-} (b-x)^k=0 . \]
The Mittag-Leffler function  $E_{\alpha}[\lambda (x-a)^{\alpha}]$
is invariant with respect to the Caputo derivatives $\ ^CD^{\alpha}_{a+}$,
\[ ^CD^{\alpha}_{a+} E_{\alpha}[\lambda (x-a)^{\alpha}]=
\lambda  E_{\alpha}[\lambda (x-a)^{\alpha}] , \]
but it is not the case for the Caputo derivative $^CD^{\alpha}_{b-}$.

\subsection{Fractional Liouville derivative}

Let us define the fractional Liouville 
derivative on the whole real axis $R$. 
The Liouville fractional derivatives $D^{\alpha}_{+}$ and $D^{\alpha}_{-}$ 
of order $\alpha >0$ are defined \cite{KST} by
\[ (D^{\alpha}_{+}f)(x)=D^n_x (I^{n-\alpha}_{+})(x)= \]
\[ = \frac{1}{\Gamma(n-\alpha)} \frac{\partial^n}{\partial x^n} 
\int^x_{-\infty} \frac{f(z)dz}{(x-z)^{\alpha-n+1}} , \]
\[ (D^{\alpha}_{-}f)(x)=(-1)^n D^n_x (I^{n-\alpha}_{-})(x)= \]
\[ = \frac{(-1)^n}{\Gamma(n-\alpha)} \frac{\partial^n}{\partial x^n} 
\int^{+\infty}_x \frac{f(z)dz}{(z-x)^{\alpha-n+1}} , \]
where $n=[\alpha]+1$ and $[\alpha]$ means the integral part of $\alpha$.
Here $D^n_x$ is the usual  derivative of order $n$. 
In particular, when $\alpha=n \in N$, then
\[ (D^0_{+}f)(x)=(D^0_{-}f)(x)=f(x) , \]
\[ (D^n_{+}f)(x)=D^n_x f(x) , \quad (D^n_{-}f)(x)=(-1)^n D^n_x f(x) . \]

If $f(x)$ is an integrable function and $\beta>\alpha>1$, then
\[ (D^{\alpha}_{\pm} I^{\alpha}_{\pm} f)(x)=f(x) , \]
\[ (D^{\alpha}_{\pm} I^{\beta}_{\pm} f)(x)=(I^{\beta-\alpha}_{\pm} f)(x) , \]
\[ (D^k D^{\alpha}_{\pm} f)(x)=(\pm 1)^k (D^{\alpha+k}_{\pm} f)(x) . \]

If $\alpha >0$, then the following relations hold:
\[ ({\cal F} D^{\alpha}_{\pm} f)(k)=(\mp i k)^{\alpha} ({\cal F}f)(k) , \]
where
\[ (\mp i k)^{\alpha}= |k|^{\alpha} \exp \{ \pm \alpha \pi i sgn(x) /2 \} .  \]
Here ${\cal F}$ is the Fourier transform.

\subsection{Interpretation of equations with fractional derivatives}

To describe the physical interpretation 
of equations with fractional derivatives and integrals with respect to time, 
we consider the memory effects and limiting cases widely used in physics: 
(1) the absence of the memory; 
(2) the complete memory;
(3) the power-like memory. 

Let us consider the evolution of a dynamical system 
in which some quantity $F(t)$ is related 
to another quantity $f(t)$ through a memory function $M(t)$: 
\be \label{4-convol} 
F(t)= \int^t_0 M(t-\tau) f(\tau) d \tau.  
\ee
Equation (\ref{4-convol}) means that
the value $F(t)$ is related with $f(t)$ by the convolution operation
\[ F(t)=M(t)*f(t) . \]
Equation (\ref{4-convol}) is a typical non-Markovian equation obtained 
in studying the systems coupled to an environment, 
with environmental degrees of freedom being averaged. 
Let us consider special cases of Eq. (\ref{4-convol}). 

(1) For a system without memory, we have the Markov processes, 
and the time dependence of the memory function is 
\be \label{4-Md} M(t-\tau)=\delta(t-\tau) , \ee
where $\delta(t-\tau)$ is the Dirac delta-function. 
The absence of the memory means that the function 
$F(t)$ is defined by $f(t)$ at the only instant $t$. 
For this limiting case, the system loses all its states 
except for one with infinitely high density. 
Using (\ref{4-convol}) and (\ref{4-Md}), we have 
\be \label{4-delta}
F (t) = \int^t_0 \delta(t-\tau) f(\tau) d\tau  = f(t) . \ee
The expression (\ref{4-delta}) corresponds to the 
process with complete absence of memory. 
This process relates all subsequent states to previous states 
through the single current state at each time $t$. 

(2) If memory effects are introduced into the system the delta-function 
turns into some function, with the time interval during 
which $f(t)$ affects on the function $F(t)$.
Let $M(t)$ be the step function 
\[ M(t-\tau)= t^{-1} , \quad (0 < \tau < t) ; \]
\[ M(t-\tau)= 0, \quad (\tau> t). \]
The factor $t^{-1}$ is chosen 
to get normalization of the memory function to unity: 
\[ \int^t_0 M(\tau) d \tau=1 . \]
Then in the evolution process the system passes 
through all states continuously without any loss. 
In this case, 
\[ F(t)=\frac{1}{t} \int^{t}_0 f(\tau) d \tau , \]
and this corresponds to complete memory. 

(3) The power-like memory function 
\be \label{4-Mpl}
M(t-\tau) =M_0 \, (t-\tau)^{\varepsilon-1} 
\ee
indicates the presence of the fractional derivative or integral. 
%%%The integral representation is equivalent to a 
%%%differential equation of the fractional order. 
Substitution of (\ref{4-Mpl}) into (\ref{4-convol}) gives
the temporal fractional integral of order $\varepsilon$:
\be \label{4-FracInt}
F(t)=\frac{\lambda}{\Gamma(\varepsilon)} 
\int^t_0 (t-\tau)^{\varepsilon-1} f(\tau) d\tau , \quad (0< \varepsilon<1) ,
\ee
where $\Gamma(\varepsilon)$ is the Gamma function, 
and $\lambda=\Gamma(\varepsilon) M_0$. 
The parameter  $\lambda$  can be regarded as the strength of 
the perturbation induced by the environment of the system.
The physical interpretation of the fractional integration
is an existence of a memory effect with power-like memory function.
The memory determines an interval $t$ during which the function
$f(\tau)$ affects on the function $F(t)$.

Equation (\ref{4-convol}) is a special type of equations
for $F(t)$ and $f(t)$, where $F$ is directly proportional to $M*f$.
In the general case, the values $F(t)$ and $f(t)$
can be related by the equation
\be \label{4-GC}
E(F,M*f)=0 ,
\ee
where $E$ is a smooth function.
Relation (\ref{4-GC}) is a fractional equation for a dynamical system.
This equation describes the memory effect. 
If $F$ is a coordinate $q(t)$, either velocity $\dot{q}(t)$, 
or acceleration $\ddot{q}(t)$, 
and $f$ is a derivative $q^{(m)}(t)$, then Eq. (\ref{4-GC})
can be considered as an equation of motion with memory.
For a power-like memory function $M(t)$, we
present (\ref{4-GC}) as a equation with fractional derivatives:
$E \Bigl( q, \dot{q}, \ddot{q}(t), \, D^{\alpha}_t q \Bigr)=0$.
As the result, we can use the fractional calculus \cite{KST}
to describe the motion of systems.

\subsection{Discrete maps with memory}

The mapping $x_{n+1}=f(x_n)$ does not have any memory, as
the value $x_{n+1}$ only depends on $x_n$. 
The introduction of memory means that the discrete value $x_{n+1}$ 
is connected with the previous values 
$x_n$, $x_{n-1}$, . . . , $x_1$. 
Particularly, any system, which is described by a discrete map, 
will have a full memory, if each state of the system 
is a simple sum of all previous states:
\be \label{ur1}
x_{n+1} =\sum^n_{k =1} f(x_k),
\ee
where $f(x)$ is a function that define the discrete map.
In  general, the expression of Eq. (\ref{ur1}) can tend  to  infinity.
Note that the  full memory exists for functions that give  a
finite sum in Eq. (\ref{ur1}).
The full memory is ideal because 
has the same action upon the next states as all the others in memory.
The map with a long-term memory can be expressed as
\be \label{ur2}
x_{n+1} =\sum^n_{k=1} V_{\alpha}(n,k) f(x_k) ,
\ee
where the weights $V_{\alpha}(n,k)$, and the parameter $\alpha$ 
characterize the non-ideal memory effects.
The  forms of the functions $f(x)$ and $V_{\alpha} (n,k)$ in Eq. (\ref{ur2}) 
are obtained by the differential equation. 
Note that linear  differential (kicked) equations 
gives the linear function $f(x)=x$.
The discrete maps with memory is considered, for example, in the papers
\cite{Ful,Fick1,Fick2,Giona,Gallas,Stan,JPA2008,JPA2009,PLA2009}. 
The interesting question is a connection of fractional equation 
of motion and the discrete maps with memory \cite{JPA2008,JPA2009}.
It is important to derive discrete maps with memory 
from equation of motion with fractional derivatives.

%%%%%%%%%%%%%%%%%%%%%%%%%%%%%%%%%%%%%%%%%%%%%%%%%%%%%%%%%%%%%%%%%%%%%%%%%%%%%%
\section{Fractional Zaslavsky map}

In this section, a brief review of discrete maps 
is considered to fix notations and provide convenient references. 
For details, see \cite{Zbook1,Zbook3,Chirikov,Schuster,CE}.
Fractional generalizations of these maps are discussed.
The fractional Zaslavsky maps are suggested.

\subsection{Discrete Chirikov and Zaslavsky maps}

Let us consider the equations of motion
\be \label{eq1}
\ddot{x}+K G(x) \sum^{\infty}_{n=0} \delta \Bigl(\frac{t}{T}-n \Bigr)=0
\ee
in which perturbation is a periodic sequence of delta-function type pulses (kicks)
following with period $T=2\pi / \nu$, $K$ is an amplitude of the pulses, 
and $G(x)$ is a some function.
This equation can be presented in the Hamiltonian form
\be \label{HE1}
\dot{x}=p , \quad
\dot{p}+K G(x) \sum^{\infty}_{n=0} \delta \Bigl(\frac{t}{T}-n \Bigr)=0 .
\ee

It is well-known that these equations can be represented in the form
of discrete map (see for example Chapter 5 in \cite{Zbook3}).
The solution of the left side of the $n$-th kick
\be \label{not1}
x_n=x(t_n-0)=\lim_{\varepsilon \rightarrow 0+} x(nT-\varepsilon), \quad 
p_n=p(t_n-0)=\lim_{\varepsilon \rightarrow 0+} p(nT-\varepsilon), \quad 
t_n=nT
\ee
is connected with the solution on the right hand side of the kick
$x(t_n+0)$, $p(t_n+0)$ by Eq. (\ref{HE1}), and 
the condition of continuity $x(t_n+0)=x(t_n-0)$. 
Using notation (\ref{not1}), 
we can derive the iteration equations
\be \label{E1}
x_{n+1}=x_n+p_{n+1}T , \ee
\be \label{E2}
p_{n+1}=p_n-KT\, G(x_n) .
\ee
Equations (\ref{E1}) and (\ref{E2}) are called the universal map.
If $T=1$ and $G(x) = \sin(x)$, then Eqs. (\ref{E1}) and (\ref{E2})
give the Chirikov map. 

%%%%%%%%%%%%%%%%%%%%%%%%%%%%%%%%%%%%%%%%%%%%%%%%%%%%%%%%%%%%%%%

The Zaslavsky map \cite{DM1,DM2,DM3} is defined by 
\be \label{Zasl1a}
X_{n+1} = X_n +  \mu Y_{n+1}+\Omega , 
\ee
\be \label{Zasl2a}
Y_{n+1} = e^{-q} [ Y_n + \varepsilon \, \sin(X_n) ] ,
\ee
where we use the parameter
\[ \mu=(e^q-1)/q. \]
Note that a shift $\Omega$ does not play an important role 
and it can be put zero ($\Omega=0$). 
The Zaslavsky map with $\Omega=0$ can be represented by the equations
\be \label{Zasl1}
X_{n+1}=X_n+P_{n+1} ,
\ee
\be \label{Zasl2}
P_{n+1}=-bP_n-Z \, \sin(X_n) .
\ee
For $b=-1$ and $Z=K$, we get the standard map 
that is described by Eqs. (\ref{E1}) and (\ref{E2}) with $T=1$. 
For the parameters
\be \label{cond} Z= - \varepsilon  \mu e^{-q} , \quad 
P_n= \mu Y_n , \quad b=-e^{-q}, \quad \mu = (e^{q}-1)/q  \ee
equations (\ref{Zasl1}) and (\ref{Zasl2}) give
Eqs. (\ref{Zasl1a}) and (\ref{Zasl2a}) with $\Omega=0$. 
For large $q \rightarrow \infty$ (for small $b \rightarrow 0$)
Eqs. (\ref{Zasl1}) and (\ref{Zasl2}) with $Z=-K$
shrink to the one-dimensional sine-map
\be \label{sine-map} 
X_{n+1}=X_{n}+K \sin (X_n)
\ee
proposed in \cite{Ar} and studied in many papers.

%%%%%%%%%%%%%%%%%%%%%%%%%%%%%%%%%%%%%%%%%%%%%%%%%%%%%%%%%%%%%%%%%%%%%%%%%%%%

\subsection{Fractional universal and Zaslavsky map}

Let us consider a fractional generalization of (\ref{eq1}) in the form
\be \label{eq2}
_0D^{\alpha}_t x+K G(x) \sum^{\infty}_{n=0} \delta \Bigl(\frac{t}{T}-n \Bigr)=0, 
\quad (1 <\alpha \le 2) ,
\ee
where $ _0D^{\alpha}_t$ is the Riemann-Liouville fractional derivative
\cite{SKM,Podlubny,KST}, which is defined by
\be \label{RLFD}
_0D^{\alpha}_t x=D^2_t \ _0I^{2-\alpha}_t x=
\frac{1}{\Gamma(2-\alpha)} \frac{d^2}{dt^2} \int^{t}_0 
\frac{x(\tau) d \tau}{(t-\tau)^{\alpha-1}} , \quad (1 <\alpha \le 2) .
\ee
Here we use the notation $D^2_t=d^2/dt^2$, and 
$ _0I^{\alpha}_t$ is a fractional integration \cite{SKM,Podlubny,KST}. 
The discrete map that corresponds to the fractional differential equation 
of order $1 <\alpha \le 2$ is derived in \cite{JPA2008,JPA2009}. 
This map can be considered as a generalization of universal map
for the case $1 <\alpha \le 2$. \\

%%%\begin{theorem}
\vskip 3mm \noindent {\bf Theorem 1}. {\it 
Fractional differential equation of kicked system (\ref{eq2})
the initial conditions
\[ ( _0D^{\alpha-1}_t x) (0+) = p_1, \quad ( _0D^{\alpha-2}_t x)(0+) = C, \]
is equivalent to the discrete map
\be 
\label{Main1}
p_{n+1}=p_n-KT \, G(x_n) ,
\ee
\be 
\label{Main2}
x_{n+1}=\frac{T^{\alpha-1}}{\Gamma(\alpha)} \sum^{n}_{k=0} p_{k+1} V_{\alpha}(n-k+1) 
+\frac{CT^{\alpha-2}}{\Gamma(\alpha-1)} (n+1)^{\alpha-2} ,
\ee
where $1<\alpha\le 2$ and the function $V_{\alpha}(z)$ is defined by
\be 
\label{V}
V_{\alpha}(z)=z^{\alpha-1}-(z-1)^{\alpha-1} \quad (z \ge 0) .
\ee
} \\ 

Proof of this Proposition is realized in \cite{JPA2008} for $C=0$, 
and in \cite{JPA2009} for $C \not=0$. 
Equations (\ref{Main1}) and (\ref{Main2}) 
define a fractional universal map in the phase space $(x_n,p_n)$,
which is a generalization of the map (\ref{E1}) and (\ref{E2}). 
Note that Eqs. (\ref{Main1}), (\ref{Main2}) with $G(x)=\sin (x)$ give
\be \label{fsm1}
p_{n+1}=p_n-KT \, \sin ( x_n ) , \quad (1<\alpha\le 2) ,
\ee
\be \label{fsm2}
x_{n+1}=\frac{T^{\alpha-1}}{\Gamma(\alpha)} \sum^{n}_{k=0} p_{k+1} V_{\alpha}(n-k+1) 
+\frac{C T^{\alpha-2}}{\Gamma(\alpha-1)} (n+1)^{\alpha-2} .
\ee
These equations define a fractional standard map on the phase space, 
which can be called the fractional Chirikov-Taylor map. 

The fractional generalization of Zaslavsky map \cite{DM1,DM2} 
can be defined by 
\be \label{Zaslmap1}
p_{n+1}=-bp_n-Z \, \sin ( x_n ) , \quad (1<\alpha\le 2) ,
\ee
\be \label{Zaslmap2}
x_{n+1}=\frac{1}{\Gamma(\alpha)} \sum^{n}_{k=0} p_{k+1} V_{\alpha}(n-k+1) 
+\frac{C}{\Gamma(\alpha-1)} (n+1)^{\alpha-2} ,
\ee
where the parameters are defined by conditions (\ref{cond}). 
For $b=-1$ and $Z=K$, Eqs. (\ref{Zaslmap1}) and (\ref{Zaslmap2}) give 
the fractional standard map that is described by 
Eqs. (\ref{fsm1}) and (\ref{fsm2}) with $T=1$. 
For for small $b \rightarrow 0$ and $C=0$,
the equations (\ref{Zaslmap1}) and (\ref{Zaslmap2}) with $Z=-K$
shrink to a one-dimensional map
\be \label{fsine-map}
x_{n+1}=  \frac{K}{\Gamma(\alpha)} 
\sum^{n}_{k=0} \, V_{\alpha}(n-k+1) \, \sin ( x_n ) , \quad (1<\alpha\le 2).
\ee
This map can be considered as a fractional generalization of 
one-dimensional sine-map (\ref{sine-map}).
The fractional sine-map (\ref{fsine-map}) is characterized by a long-term memory
such that the present state evolution depends 
on all past states with special forms of weights function (\ref{V}).

%%%%%%%%%%%%%%%%%%%%%%%%%%%%%%%%%%%%%%%%%%%%%%%%%%%%%%%%%%%%%%%%%%%%%%%%%%%%%%%%%%%

\subsection{Kicked damped rotator map}

Let us consider a kicked damped rotator \cite{Schuster}. 
The equation of motion for this rotator is
\be \label{dr}
\ddot{x}+ q \dot{x}=K G(x) \sum^{\infty}_{n=0} \delta \Bigl(t -n T \Bigr) .
\ee
It is well-known that Eq. (\ref{dr}) gives \cite{Schuster} 
the two-dimensional map
\be \label{dr-map1}
p_{n+1}=e^{-qT} [ p_n+K G(x_n)] ,
\ee
\be \label{dr-map2}
x_{n+1}=x_n+\frac{1-e^{-qT}}{q} [ p_n+K G(x_n)] .
\ee
This map is known as the kicked damped rotator map. 
The phase volume shrinks each time step by a factor $\exp(-q)$. 
The map is defined by two important parameters, 
dissipation constant $q$ and force amplitude $K$. 
These equations can be rewritten in the form
\be 
p_{n+1}=e^{-qT} [ p_n+K G(x_n)] ,
\ee
\be 
x_{n+1}=x_n+\frac{e^{qT}-1}{q} p_{n+1} .
\ee
It is easy to see that these equations give
the Zaslavsky map (\ref{Zasl1}) and (\ref{Zasl2}) with $\Omega=0$ 
if 
\[ X_n=x_n, \quad Y_n=p_n, \quad \varepsilon=K, \quad T=1, \quad G(x)=\sin(x). \]
If we consider the limit $q \rightarrow \infty$ and $K \rightarrow \infty$ 
such that $q/K=1$ and the function
\[ G(x)=(r-1)x-r x^2 , \]
then equations (\ref{dr-map2}) and (\ref{dr-map2}) give the logistic map
\[ x_{n+1}=r x_n (1-x_n) . \]
Note that the evolution of logistic maps with 
a long-term memory is considered in \cite{Stan}.

\subsection{Fractional Zaslavsky map from fractional differential equations}

Let us consider a fractional generalization of 
differential equation for a kicked damped rotator (\ref{dr}) in the form 
\be \label{fdr}
_0D^{\alpha}_t x - q \, _0D^{\beta}_t x =
K G(x) \sum^{\infty}_{n=0} \delta (t -n T ) ,
\ee
where $q\in R$ and $ _0D^{\alpha}_t$ is derivative of non-integer order.
This equation will be considered with
\[ 1< \alpha\le 2 , \quad \beta=\alpha-1 , \]
and $ _0D^{\alpha}_t$ is the Riemann-Liouville fractional derivative
\cite{SKM,Podlubny,KST} defined by (\ref{RLFD}). 
Note that we use the minus in the left hand side of Eq. (\ref{fdr}), 
where $q$ can be positive and negative value. 
The discrete map that corresponds to the fractional 
differential equation has been derived in \cite{JPA2008}. \\

\vskip 3mm \noindent {\bf Theorem 2}. {\it 
Fractional differential equation of kicked system (\ref{fdr})
is equivalent to the discrete map
\be 
\label{New1}
p_{n+1}=e^{qT} \Bigl[ p_n+K G ( x_n ) \Bigr] ,
\ee
\be 
\label{New2}
x_{n+1}=\frac{1}{\Gamma(\alpha-1)} 
\sum^{n}_{k=0} p_{k+1} \, W_{\alpha}(q,T,n+1-k) ,
\ee
where the functions $W_{\alpha}(q,T,z)$ are defined by 
\be 
\label{Wa}
W_{\alpha}(q,T,m+1) = T^{\alpha-1} 
\int^{1}_0 \, e^{-qTy} \, (m+y)^{\alpha-2} \, d y .
\ee
} \\

Proof of this statement is realized in \cite{JPA2008}.
Equations (\ref{New1}) and (\ref{New2}) 
define a fractional kicked damped rotator map.

This Theorem can be used to derive 
a fractional generalization of the Zaslavsky map.
If we use the conditions
\be \label{Zcond}
X_n=x_n, \quad Y_n=p_n,  \quad \varepsilon=K, \quad T=1, \quad G(x)=\sin(x) ,
\ee
%%%b= - \exp \{q \} , \quad T=1, \quad e^{q} K = - Z , \quad G(x)=\sin (x) , \ee
then Eqs. (\ref{New1}) and (\ref{New2}) give the map
\be \label{New1z}
Y_{n+1}=e^{q} \Bigl[ Y_n+ \varepsilon \sin ( X_n ) \Bigr] ,
\ee
\be \label{New2z} 
X_{n+1}=\frac{1}{\Gamma(\alpha-1)} 
\sum^{n}_{k=0} Y_{k+1} \, W_{\alpha}(q,1,n+1-k)   .
\ee
These equations can be considered as a second fractional generalization 
of the Zaslavsky map (\ref{Zasl1a}) and (\ref{Zasl2a}) with $\Omega=0$.
Note that this fractional Zaslavsky map is derived from fractional 
nonlinear differential equation (\ref{fdr}) with condition (\ref{Zcond}).
If we use the variables
\[ P_n= \mu Y_n , \quad b = - e^q , \quad Z = - \mu \varepsilon e^q , \]
then discrete equations 
\be 
P_{n+1}= - b P_n -Z \sin ( X_n ) ,
\ee
\be 
X_{n+1} = \frac{\mu^{-1}}{\Gamma(\alpha-1)} 
\sum^{n}_{k=0} P_{k+1} \, W_{\alpha}(q,1,n+1-k) .
\ee
These equations can be considered as a fractional generalization 
of the Zaslavsky map  equations (\ref{Zasl1}) and (\ref{Zasl2}) with $\Omega=0$.
For $\alpha=2$, this fractional Zaslavsky map gives the Zaslavsky map 
that is described by Eqs. (\ref{Zasl1}), and (\ref{Zasl2}).
For $\exp \{q \} \rightarrow 0$ 
equations (\ref{New1z}) and (\ref{New1z}) 
shrink to a one-dimensional map that 
can be considered as a fractional generalization of 
one-dimensional sine-map (\ref{sine-map}).

%%%%%%%%%%%%%%%%%%%%%%%%%%%%%%%%%%%%%%%%%%%%%%%%%%%%%%%%%%%%%%%%%%%%%%%%%%%%%%%%%%%%%%%%%

\section{Fractional H\'enon map}

\subsection{H\'enon map}

The H\'enon map \cite{Henon,RHO} can be considered as a two-dimensional 
generalization of the logistic map \cite{Schuster}.
Then the H\'enon map is presented as a map given by the coupled equations
\be \label{Hen1}
x_{n+1}=1-ax^2_n+y_n ,
\ee
\be \label{Hen2}
y_{n+1}=bx_n .
\ee
Substitution of Eq. (\ref{Hen2}) into Eq. (\ref{Hen1}) gives
\be \label{Hen3}
x_{n+1}= 1 - ax^2_n + bx_{n-1} .
\ee
This equation is equivalent to the H\'enon map that is defined by 
Eqs. (\ref{Hen1}) and (\ref{Hen2}).
Note that the H\'enon map in the form (\ref{Hen3}) 
can be derived from the generalized dissipative map.

%%%%%%%%%%%%%%%%%

Let us consider the generalized dissipative map
\be \label{D1} 
x_{n+1}=x_n +p_{n+1}T, 
\ee
\be \label{D2}
p_{n+1}=-bp_{n}+K T G(x_n) , \quad |b|\le 1 .
\ee
Using Eqs. (\ref{D1}) and (\ref{D2}) with
$K=T=1$, and the function 
\be \label{Ghen}
G (x)= 1 - (1-b) x- ax^2 ,
\ee
the H\'enon map in the form (\ref{Hen3}) can be obtained.
Let us prove this statement. 
Substitution of (\ref{D1}) in the form 
$p_{n+1}=x_{n+1}-x_n$ into (\ref{D2}) gives
\be
x_{n+1}-x_n=-b(x_n-x_{n-1})+ G(x_n) .
\ee
Then we have
\be \label{Hen4} 
x_{n+1}-bx_{n-1} = (1-b)x_n + G(x_n) . 
\ee
Substitution of (\ref{Ghen}) into (\ref{Hen4}) gives the H\'enon map (\ref{Hen3}). 
%%%\be \label{Hen5} x_{n+1}-bx_{n-1} =  1 - ax^2  . \ee

%%%%%%%%%%%%%%%

\subsection{Fractional H\'enon map}

To derive a fractional generalization of the H\'enon map
we can use the fractional generalized dissipative map and 
the fact that the H\'enon map in the form (\ref{Hen3}) 
can be derived from the generalized dissipative map.

Let us consider a fractional generalization of the 
map (\ref{D1}) and (\ref{D2}) in the form
\be \label{fgdm1}
x_{n+1}= \frac{T^{\alpha-1}}{\Gamma(\alpha)} 
\sum^{n}_{k=0} p_{k+1} V_{\alpha}(n-k+1), \quad (1<\alpha\le 2) ,
\ee
\be \label{fgdm2}
p_{n+1}=-bp_n+K T \, G ( x_n ) .
\ee
This is the fractional generalized dissipative map. 
If $\alpha=2$, then Eqs. (\ref{fgdm1}) and (\ref{fgdm2}) 
give (\ref{D1}) and (\ref{D2}).
If $G(x)=\sin(x)$, $T=1$ and $K=-Z$, $C=0$, then 
Eqs. (\ref{fgdm1}) and (\ref{fgdm2}) 
give the fractional Zaslavsky map equations 
(\ref{Zaslmap1}) and (\ref{Zaslmap2}).

To derive a fractional generalization of the H\'enon map, 
we can use the fractional dissipative map (\ref{fgdm1}) and (\ref{fgdm2})
with $K=T=1$, and the function (\ref{Ghen}).
As a result, we obtain the equations
\be \label{FDM1}
x_{n+1}=\frac{1}{\Gamma(\alpha)} \sum^{n}_{k=0} p_{k+1} V_{\alpha}(n-k+1) ,
\ee 
\be \label{FDM2}
p_{n+1}= - b p_n + 1 - (1-b)x_n - a x^2_n ,\quad (|b|\le 1) ,
\ee
where $ 1<\alpha\le 2 $ and $V_{\alpha}(z)$ is defined by (\ref{V}). 
%%%For $b=-1$, we get the fractional universal map (\ref{FSM}). 
These equations can be considered as a first fractional generalization of 
the H\'enon map (\ref{Hen1}), (\ref{Hen2}).

%%%%%%%%%%%%%%%%%%%%%%%%%%%%%%%%%%%%%%%%%%%%%%%%%%%%%%%%%%%%%%%%%%%%%%%%%%%%
%%%\section{Fractional H\'enon map from FKDR map}

Fractional H\'enon map (\ref{FDM1}) and (\ref{FDM2}) 
is derived from Eq. (\ref{fgdm1}) and (\ref{fgdm2}),
where the dissipation has been introduced into the iteration equation.
In this case the fractional equation of motion with dissipation is not used.
Equations (\ref{FDM1}) and (\ref{FDM2}) are not derived from
fractional differential equations of kicked damped system.
We can suggest the second possible getting of 
a fractional generalization of the H\'enon map.
This map can be derived from the fractional equation 
of kicked damped rotator (\ref{fdr}).

It is known (see Sec. 1.2 of \cite{Schuster}) that the H\'enon map 
can be derived from equation of motion (\ref{dr}) with
\be \label{Ghenon}
G (x)= -\frac{q}{1+b} \left[ 1 + (1+b) x+ ax^2 \right] ,
\ee
where $b=-\exp(-q)$ and $T=1$.
Using Eq. (\ref{dr}) with function (\ref{Ghenon}), we obtain the map equation
\be \label{dr-map1b}
x_{n+1}=x_n+\frac{1+b}{q} p_n - 1 - (1+b) x_n- ax^2_n ,
\ee
\be \label{dr-map2b}
p_{n+1}=e^{-q} p_n + \frac{qb}{1+b} \left[ 1 + (1+b) x_n+ ax^2_n \right] ,
\ee
where $T=1$, and $b=- \exp\{-q\}$.
These equations are equivalent 
to the H\'enon map equation (\ref{Hen3}).
Proof of this proposition is realized in \cite{Schuster}.

Let us consider the fractional differential equation (\ref{fdr}) 
that are generalization of Eq. (\ref{dr}) 
by non-integer order derivatives.
In order to derive fractional generalization of 
H\'enon map, we use Eq. (\ref{fdr}) 
with the function (\ref{Ghenon}).
As a result, we obtain the equations
\be \label{New1-h}
x_{n+1}=\frac{1}{\Gamma(\alpha-1)} 
\sum^{n}_{k=0} p_{k+1} \, W_{\alpha}(q,T,n+1-k) ,
\ee
\be \label{New2-h}
p_{n+1} = e^{q} \Bigl[ p_n+ 
\frac{q}{1+b} \left( 1 + (1-b) x_n+ ax^2_n \right) \Bigr] ,
\ee
where the functions $W_{\alpha}(q,T,z)$ 
are defined by Eq. (\ref{Wa}) and $q=\ln(-b)$. 
These equations define a second fractional generalization of 
the H\'enon map.
This map is derived from nonlinear differential equations with 
derivatives of non-integer order.
Equations (\ref{New1-h}) and (\ref{New2-h}) can be called fractional
H\'enon map.

%%%%%%%%%%%%%%%%%%%%%%%%%%%%%%%%%%%%%%%%%%%%%%%%%%%%%%%%%%%%%%%%%%%%%%%%%%%%

\section{Fractional derivative in the kicked term and Zaslavsky map}

In this section, a fractional generalization of 
differential equation (\ref{eq1}) is suggested. 
In this generalization, we use a fractional derivative of 
the order $0 \le \beta < 1$ in the kicked term, i.e. the term 
of a periodic sequence of delta-function type pulses (kicks).
The other generalization that is described by Eq. (\ref{eq2}) 
is suggested in \cite{JPA2008}.
The discrete map that corresponds to the suggested fractional equation 
of order $0 \le \beta < 1$ is derived. 
This map can be considered as a generalization of universal map
for the case $0 <\beta < 1$.

\subsection{Fractional equation and discrete map}

Let us consider a fractional generalization of (\ref{eq1}) in the form
\be \label{New1zz}
D^2_t x +K G(\, _0^CD^{\beta}_t x ) 
\sum^{\infty}_{n=0} \delta \Bigl(\frac{t}{T}-n \Bigr)=0, 
\quad (0 \le \beta <1) .
\ee
where $ _0^CD^{\beta}_t$ is the Caputo fractional derivative
\be \label{CFD}
_0^CD^{\beta}_t x=\ _0I^{1-\beta}_t D^1_t x=
\frac{1}{\Gamma(1-\beta)} \int^{t}_0 
\frac{d \tau}{(t-\tau)^{\beta}} \frac{d x(\tau)}{d \tau} ,
\quad (0 \le \beta<1).
\ee
Here we use the notation $D^1_t=d/dt$, and 
$ _0I^{\alpha}_t$ is a fractional integration \cite{SKM,Podlubny,KST}. 
For $\beta=0$ fractional equation (\ref{New1zz}) gives Eq. (\ref{eq1}). \\

%%%\begin{theorem}
\vskip 3mm \noindent {\bf Theorem 3}. {\it 
Fractional differential equation of kicked system (\ref{New1zz})
is equivalent to the discrete map
\be \label{First-n}
x_{n+1}=x_n+p_{n+1}T ,
\ee
\be \label{F3-n}
p_{n+1}=p_n-KT \,
G\Bigl( \frac{T^{1-\beta}}{\Gamma(2-\beta)} 
\sum^{n-1}_{k=0} p_{k+1} V_{2-\beta}(n-k) \Bigr) ,
\quad (0\le \beta <1),
\ee
where the function $V_{2-\beta}(z)$ is defined by
\be 
%%%\label{V}
V_{2-\beta}(z)=z^{1-\beta}-(z-1)^{1-\beta} \quad (z \ge 1).
\ee
}\\

\bp
The fractional equation (\ref{New1zz}) can be presented in the Hamiltonian form
\[
\dot{x}=p ,
\]
\be \label{HE2-n}
\dot{p}+K G(\, _0^CD^{\beta}_t x ) 
\sum^{\infty}_{n=0} \delta \Bigl(\frac{t}{T}-n \Bigr)=0 .
\ee

Between any two kicks there is a free motion
\be \label{sol2-n}
p=const, \quad x=p t+const . \ee
The integration of (\ref{HE2-n}) over $(t_n+\varepsilon,t_{n+1}-\varepsilon)$ gives
\be \label{z1-n1}
x(t_{n+1}-0)=x(t_n+0)-p_{n+1} T , 
\ee
\be \label{z1-n2}
p(t_{n+1}-0)=p(t_n+0) .
\ee

The solution of the left side of the $n$-th kick
\be \label{not2-n1}
x_n=x(t_n-0)=\lim_{\varepsilon \rightarrow 0} x(nT-\varepsilon), 
\ee
\be \label{not2-n2}
p_n=p(t_n-0)=\lim_{\varepsilon \rightarrow 0} p(nT-\varepsilon), \quad 
t_n=nT,
\ee
is connected with the solution on the right hand side of the kick
$x(t_n+0)$, $p(t_n+0)$ by Eq. (\ref{HE2-n}), and 
the continuity condition
\be \label{z2-n} 
x(t_n+0)=x(t_n-0) . 
\ee
The integration of (\ref{HE2-n}) over the interval 
$(t_n-\varepsilon,t_n+\varepsilon)$ gives
\be \label{z3-n}
p(t_n+0)=p(t_n-0)-KT\, G( \, _0^CD^{\beta}_{t_n} x ) . 
\ee
Using notations (\ref{not2-n1}) and (\ref{not2-n2}), 
and the solution (\ref{sol2-n}), we get
\be
p(t_n+0)=p(t_{n+1}-0)=p_{n+1} .
\ee
Substituting of (\ref{z2-n}) and (\ref{z3-n}) into (\ref{z1-n1}) and (\ref{z1-n2}), 
we derive the iteration equations
\be
x_{n+1}=x_n+p_{n+1}T ,
\ee
\be \label{F2-n}
p_{n+1}=p_n-KT\, G(\, _0^CD^{\beta}_{t_n} x ) .
\ee

To derive a map, we should express the fractional derivative 
\be
_0^CD^{\beta}_{t_n} x= \frac{1}{\Gamma(1-\beta)} \int^{t_n}_0 
\frac{d \tau}{(t_n-\tau)^{\beta}} \frac{d x(\tau)}{d \tau} 
\ee
through the variables (\ref{not2-n1}) and (\ref{not2-n2}). 
Using $\dot{x}=p$, we have
\be
_0^CD^{\beta}_{t_n} x=
\frac{1}{\Gamma(1-\beta)} \int^{t_n}_0 
\frac{p(\tau) d \tau}{(t_n-\tau)^{\beta}} .
\ee
It can be presented as
\be \label{tk-n}
_0^CD^{\beta}_{t_n} x= \frac{1}{\Gamma(1-\beta)} 
\sum^{n-1}_{k=0} \int^{t_{k+1}}_{t_k} 
\frac{ p(\tau) d \tau}{(t_n-\tau)^{\beta}} ,
\ee
where $t_{k+1}=t_k+T=(k+1)T$, $t_k=kT$, and $t_0=0$.

For the interval $(t_k,t_{k+1})$, Eqs. (\ref{sol2-n}), 
(\ref{not2-n1}) and (\ref{not2-n2}) give
\be
p(\tau)=p(t_k+0)=p(t_{k+1}-0)=p_{k+1}, \quad \tau \in (t_k,t_{k+1}) . 
\ee
Then 
\[ \int^{t_{k+1}}_{t_k} \frac{ p(\tau) d \tau}{(t_n-\tau)^{\beta}} =
p_{k+1} \int^{t_{k+1}}_{t_k} (t_n-\tau)^{-\beta} d\tau = \]
\[  =p_{k+1} \int^{t_n-t_k}_{t_n-t_{k+1}} z^{-\beta} dz =
p_{k+1} \frac{z^{1-\beta}}{1-\beta} \Bigl|^{t_n-t_k}_{t_n-t_{k+1}}= \]
\[ = p_{k+1} \frac{1}{1-\beta} 
\Bigl[(t_n-t_k)^{1-\beta}-(t_n-t_{k+1})^{1-\beta} \Bigr]= \]

\be \label{tk2-n}
=p_{k+1} \frac{T^{1-\beta}}{1-\beta} 
\Bigl[ (n-k)^{1-\beta}-(n-k-1)^{1-\beta} \Bigr]. 
\ee
Using $(1-\beta)\Gamma(1-\beta)=\Gamma(2-\beta)$ and Eq. (\ref{tk2-n}), 
the fractional derivative (\ref{tk-n}) can be presented as
\be \label{29-n}
_0^CD^{\beta}_{t_n} x=
\frac{T^{1-\beta}}{\Gamma(2-\beta)} \sum^{n-1}_{k=0} p_{k+1} V_{2-\beta}(n-k) ,
\quad (0 \le \beta <1),
\ee
where
\be \label{V-n}
V_{2-\beta}(z)=z^{1-\beta}-(z-1)^{1-\beta} .
\ee

This ends of the proof. 
\ep

As a result, Eq. (\ref{F2-n}) takes the form of Eqs. (\ref{First-n}) and (\ref{F3-n}). 
These equations define the fractional generalization of universal map. 
Let us consider the following Corollary. \\

%%%\begin{theorem} 
\vskip 3mm \noindent {\bf Theorem 4}. {\it 
For $\beta=0$ the fractional universal map, 
which is defined by Eqs. (\ref{First-n}) and (\ref{F3-n}), 
gives the usual universal map.} \\

\bp
Let us consider that the fractional universal map 
(\ref{First-n}) and (\ref{F3-n}) for $\beta=0$. 
Substitution of Eq. (\ref{First-n}) in the form
\be \label{eq3-n}
p_{k+1}=\frac{1}{T}(x_{k+1}-x_k) 
\ee
into iteration equation (\ref{F3-n}) gives
\be \label{xixi-n}
p_{n+1}=p_n-KT \,
G\Bigl( \frac{T^{-\beta}}{\Gamma(2-\beta)} 
\sum^{n-1}_{k=0} (x_{k+1}-x_{k}) V_{2-\beta }(n-k) \Bigr) .
\ee
Then, the fractional map (\ref{First-n}), (\ref{F3-n}) is defined by 
\[ x_{n+1}=x_n+p_{n+1}T , \]
\be \label{FUM2-n}
p_{n+1}=p_n-KT \,
G\Bigl( \frac{T^{-\beta}}{\Gamma(2-\beta)} 
\sum^{n}_{k=1} (x_{k}-x_{k-1}) V_{2-\beta }(n-k) \Bigr) . 
\ee
where $0\le \beta <1$.
For $\beta=0$, we have $V_{2-\beta}(z)=1$, and
\be 
\sum^{n-1}_{k=0} (x_{k+1}-x_{k}) = x_n-x_0 .
\ee 
Then Eq. (\ref{xixi-n}) gives
\be
p_{n+1}=p_n-KT \, G \Bigl( x_n-x_0 \Bigr) .
\ee
As a result, Eqs. (\ref{FUM2-n}) with $\beta=0$ give 
the usual map for the case $x_0=0$. 

This ends of the proof. 
\ep

%%%%%%%%%%%%%%%%%%%%%%%%%%%%%%%%%%%%%%%%%%%%%%%%%%%%%%%%%%%%%%%%%%%%%%%%

\subsection{Examples}

Let us consider some examples of 
the fractional map (\ref{First-n}), (\ref{F3-n}). \\

{\bf Example 1.}
If $G(x)=-x$, then Eqs. (\ref{First-n}), (\ref{F3-n}) are
\[ x_{n+1}=x_n+p_{n+1}T , \]
\be \label{FAM-n}
p_{n+1}=p_n+K 
\frac{T^{2-\beta}}{\Gamma(2-\beta)} \sum^{n-1}_{k=0} p_{k+1} V_{2-\beta}(n-k) ,
\quad (0 \le \beta < 1),
\ee
where $V_{2-\beta}(z)$ is defined in (\ref{V}).
Equations (\ref{FAM-n}) can be considered as a fractional 
generalization of the Anosov type system.  \\

{\bf Example 2.}
If $G(x)=\sin (x)$, Eqs. (\ref{First-n}), (\ref{F3-n}) give
\[ 
x_{n+1}=x_n+p_{n+1}T ,
\]
\be \label{F4-n}
p_{n+1}=p_n+KT\, 
\sin \Bigl( \frac{T^{1-\beta}}{\Gamma(2-\beta)} 
\sum^{n-1}_{k=0} p_{k+1} V_{2-\beta}(n-k) \Bigr) , \quad (0 \le \beta < 1) . 
\ee
This map can be considered as a fractional generalization of standard map. 
%%%which can be called the fractional Chirikov-Taylor map.
The other possible form of Eq. (\ref{F4-n}) is 
\[
x_{n+1}=x_n+p_{n+1} ,
\]
\be \label{FSM-n}
p_{n+1}=p_n-K \,
\sin \Bigl( \frac{1}{\Gamma(2-\beta)} 
\sum^{n-1}_{k=0} (x_{k+1}-x_{k}) \Bigl[(n-k)^{1-\beta}-(n-k-1)^{1-\beta} \Bigr] \Bigr) ,
\ee
where we use Eqs. (\ref{eq3-n}), (\ref{V}), $T=1$, and $0 \le \beta < 1$.
These equations define a fractional standard map that 
can be called the fractional Chirikov map. \\

{\bf Example 3.}
The fractional generalization of Zaslavsky map \cite{DM1,DM2} 
can be defined by 
\be 
x_{n+1}=x_n+p_{n+1} ,
\ee
\[ p_{n+1}= -bp_n- \]
\be \label{Zsin} 
- K \, \sin \Bigl( \sum^{n-1}_{k=0} \frac{(x_{k+1}-x_{k}) }{\Gamma(2-\beta)} 
\Bigl[(n-k)^{1-\beta}-(n-k-1)^{1-\beta}\Bigr] \Bigr) . 
\ee
For $b=1$, we get the fractional standard map (\ref{FSM-n}). 
This map is one of possible fractional generalizations of the Zaslavsky map.
In this generalization we introduce a dissipation 
by the change of the variable $p_n \rightarrow -bp_n$.
Equation (\ref{Zsin}) is not directly connected 
with a fractional equation of motion.
A generalization of the Zaslavsky map that is derived from 
the differential equation with fractional damped kicks 
is suggested in the next section.

%%%%%%%%%%%%%%%%%%%%%%%%%%%%%%%%%%%%%%%%%%%%%%%%%%%%%%%%%%%%%%%%%%%%%%%%%%%%%%

%%%%%%%%%%%%%%%%%%%%%%%%%%%%%%%%%%%%%%%%%%%%%%%%%%%%%%%%%%%%%%%%%%%%%%%%%%%%%%

\section{Fractional derivative in the kicked damped term 
and generalizations of the Zaslavsky and H\'enon maps}

In this section, a fractional generalization of 
differential equation (\ref{dr}) for a kicked damped rotator is suggested. 
In this generalization, we use a fractional derivative in 
the kicked damped term, i.e. the term 
of a periodic sequence of delta-function type pulses (kicks).
The other generalization, which is described by Eq. (\ref{fdr}), 
is suggested in \cite{JPA2008}.
The discrete map that corresponds to the suggested fractional 
differential equation is derived. 

\subsection{Fractional equation and discrete map}

Let us consider the fractional generalization of equation (\ref{dr}) in the form 
\be \label{eq7-n}
D^2_t x - q D^1_t x=
K G(\, _0^CD^{\beta}_t x ) \sum^{\infty}_{n=0} \delta (t-nT), 
\quad (0 \le \beta <1 ) ,
\ee
where $ q \in R$, 
and $ _0^CD^{\beta}_t$ is the Caputo fractional derivative 
\cite{KST} of the order $0 \le \beta<1$ defined by (\ref{CFD}). 
Note that we use the minus in the left hand side of Eq. (\ref{fdr}), 
where $q$ can be positive and negative value. 
The fractional deribative $ _0^CD^{\beta}_t x$ 
is presented in the kicked damped term. \\

%%%\begin{theorem}
\vskip 3mm \noindent {\bf Theorem 5}. {\it 
Fractional differential equation of kicked system (\ref{eq7-n})
is equivalent to the discrete map
\be \label{FDR3-n}
x_{n+1}=x_n+\frac{1-e^{-qT}}{q} p_{n+1} ,
\ee
\be \label{FDR4-n}
p_{n+1}=e^{qT} \Bigl[ p_n+K 
G \Bigl(\frac{1}{\Gamma(1-\beta)} 
\sum^{n-1}_{k=0} p_{k+1} W_{2-\beta}(q,T,n-k) \Bigr) \Bigr] ,
\ee
where the functions $W_{2-\beta}$ are defined by 
\be \label{Wa-b}
W_{2-\beta}(q,T,m) = T^{1-\beta} 
\int^{1}_0 \, e^{qT(z-1)} \, (m-z)^{-\beta} \,d z .
\ee
}\\

\bp
Fractional equation (\ref{eq7-n}) can be presented in the Hamiltonian form
\[
\dot{x}=p ,
\]
\be \label{HE3-n}
\dot{p}-q p=
K G(\, _0^CD^{\beta}_t x ) \sum^{\infty}_{n=0} \delta (t-nT), 
\quad (0 <\beta< 0, \quad q \in \mathbb{R}) .
\ee
Between any two kicks 
\be \label{67-n}
\dot{p}-q p=0 . 
\ee
For $t\in(t_n+0,t_{n+1}-0)$, the solution of Eq. (\ref{67-n}) is
\be \label{eq4-n}
p(t_{n+1}-0)=p(t_n+0) e^{qT} .
\ee
Let us use the notations $t_n=nT$, and
\[ x_n=x(t_n-0)=\lim_{\varepsilon \rightarrow 0} x(nT-\varepsilon), \]
\be \label{not3-n}
p_n=p(t_n-0)=\lim_{\varepsilon \rightarrow 0} p(nT-\varepsilon) .
\ee

For $t\in (t_n-\varepsilon,t_{n+1}-\varepsilon)$, 
the general solution of (\ref{HE3-n}) is
\be
p(t)=p_n e^{q(t-t_n)}+ 
K \sum^{\infty}_{m=0} G(\, _0^CD^{\beta}_{t_m} x )  
\int^t_{t_n-\varepsilon} d \tau e^{q(t-\tau)} \delta (\tau-mT) .
\ee
Then
\be \label{eq5-n}
p_{n+1}= e^{q T} \Bigl[ p_n+ K G(\, _0^CD^{\beta}_{t_n} x ) \Bigr] .
\ee
Using (\ref{eq5-n}), the integration of the first equation of (\ref{HE3-n}) gives
\be \label{eq6-n}
x_{n+1}=x_n - \frac{1-e^{q T}}{q} 
\Bigl[ p_n+ K G(\, _0^CD^{\beta}_{t_n} x ) \Bigr] .
\ee

Let us consider the Caputo fractional derivative 
from Eqs. (\ref{eq5-n}), (\ref{eq6-n}). Then
\[ _0^CD^{\beta}_{t_n} x= _0I^{1-\beta}_t D^1_t x=
\frac{1}{\Gamma(1-\beta)} \int^{t_n}_0 
\frac{d \tau}{(t_n-\tau)^{\beta}} \frac{d x(\tau)}{d \tau} , 
\quad (0\le \beta <1). \]
Using $\dot{x}=p$, we have
\be
_0^CD^{\beta}_{t_n} x= _0I^{1-\beta}_t p
\frac{1}{\Gamma(1-\beta)} \int^{t_n}_0 
\frac{p(\tau) d \tau}{(t_n-\tau)^{1-\beta}} .
\ee
This relation can be rewritten as
\be \label{tk3-n}
_0^CD^{\beta}_{t_n} x= \frac{1}{\Gamma(1-\beta)} 
\sum^{n-1}_{k=0} \int^{t_{k+1}}_{t_k} 
\frac{ p(\tau) d \tau}{(t_n-\tau)^{\beta}} ,
\ee
where $t_{k+1}=t_k+T=(k+1)T$, and $t_k=kT$, such that $t_0=0$.

For $\tau \in (t_k,t_{k+1})$, Eqs. (\ref{eq4-n}) and (\ref{not3-n}) give
\[ p(\tau)=p(t_k+0) e^{q(\tau-t_k)}=
p(t_{k+1}-0) e^{-qT} e^{q(\tau-t_k)}= \]
\[ =p_{k+1} e^{q(\tau-t_k-T)}=
p_{k+1} e^{q(\tau-t_{k+1})} . \]
Then 
\[ 
\int^{t_{k+1}}_{t_k} \frac{ p(\tau) d \tau}{(t_n-\tau)^{\beta}} =
p_{k+1} \int^{t_{k+1}}_{t_k} 
e^{q(\tau-t_{k+1})} (t_n-\tau)^{-\beta} d\tau =
\]

\[ 
=p_{k+1} \int^{t_n-t_k}_{t_n-t_{k+1}} 
e^{q(t_n-t_{k+1}-z)} z^{-\beta} dz =
p_{k+1} e^{q(t_n-t_{k+1})}
\int^{t_n-t_k}_{t_n-t_{k+1}} z^{-\beta} e^{-qz} dz =
\]

\be \label{y-n}
= p_{k+1} q^{\beta-1}  e^{q(n-k-1)T}
\int^{q(t_n-t_k)}_{q(t_n-t_{k+1})} y^{-\beta} e^{-y} dy .
\ee

As a result, Eq. (\ref{y-n}) can be represented by
\[ \int^{t_{k+1}}_{t_k} \frac{ p(\tau) d \tau}{(t_n-\tau)^{\beta}} = \]
\be \label{p11-n}
=p_{k+1} q^{\beta-1}  e^{q(n-k-1)T}
\Bigl[ \Gamma(1-\beta, q(t_n-t_{k+1}) ) - \Gamma(1-\beta, q(t_n-t_k) ) \Bigr] .
\ee
Here $\Gamma(a,b)$ is the incomplete Gamma function, 
where $a,b$ are complex numbers. 
This function can be defined by
\[ \Gamma(a,b)=\Gamma(a)-\frac{b^a}{a} \ _1F_1 (1,1+a;-b) . \]
Here $ _1F_1$ is the confluent hypergeometric Kummer function \cite{Erd},
%%%which can be defined by the series
\[ _1F_1(a,c;z)=\sum^{\infty}_{k=0} \frac{(a)_k}{(c)_k} \frac{z^k}{k!} , \]
and $(a)_k$ is the Pochhammer symbol
\[ (a)_k=a(a+1)...(a+k-1) , \quad k \in N . \]

Using (\ref{tk3-n}) and (\ref{p11-n}), we obtain
\be \label{tk4-n}
_0^CD^{\beta}_{t_n} x= 
\frac{1}{\Gamma(1-\beta)} \sum^{n-1}_{k=0} p_{k+1} W_{2-\beta}(q,T,n-k) ,
\quad (0 \le \beta <1) ,
\ee
where $W_{2-\beta}(q,T,m)$ is defined by (\ref{Wa-b}).

Substitution of (\ref{tk4-n}) into (\ref{eq5-n}) and (\ref{eq6-n}) gives
\be \label{FDR1-n}
p_{n+1}=e^{qT} \Bigl[ p_n+K 
G \Bigl(\frac{1}{\Gamma(1-\beta)} 
\sum^{n-1}_{k=0} p_{k+1} W_{2-\beta}(q,T,n-k)
 \Bigr) \Bigr] ,
\ee
\be \label{FDR2-n}
x_{n+1}=x_n-\frac{1-e^{qT}}{q} \Bigl[p_n+
K G \Bigl(\frac{1}{\Gamma(1-\beta)} 
\sum^{n-1}_{k=0} p_{k+1} W_{2-\beta}(q,T,n-k) \Bigr) \Bigr] .
\ee
Equations (\ref{FDR1-n}) and (\ref{FDR2-n}) can be presented  
in the form of Eqs. (\ref{FDR3-n}) and (\ref{FDR3-n}).

This ends of the proof. 
\ep

\subsection{Fractional Zaslavsky and H\'enon maps}

The iteration equations (\ref{FDR3-n}) and (\ref{FDR4-n})
define a fractional generalization of the kicked damped rotator map (\ref{dr}).
If we use the conditions (\ref{Zcond}),
%%%y_n=p_n,  \quad \varepsilon =K, \quad T=1, \quad G(x)=\sin(x). 
then Eqs. (\ref{FDR3-n}) and (\ref{FDR4-n}) give the map
\be \label{FDR3-n2}
X_{n+1}=X_n+ \mu^{\prime} Y_{n+1} ,
\ee
\be \label{FDR4-n2}
Y_{n+1}=e^{q} \Bigl[ Y_n+ \varepsilon  
\sin \Bigl(\frac{1}{\Gamma(1-\beta)} 
\sum^{n-1}_{k=0} Y_{k+1}  W_{2-\beta}(q,T,n-k) \Bigr) \Bigr] ,
\ee
where
\[ \mu^{\prime} = \frac{1-e^{-q}}{q} . \]
These equations can be considered as a fractional generalization 
of the Zaslavsky map (\ref{Zasl1a}) and (\ref{Zasl2a}) 
with $\Omega=0$ and $\mu =\mu^{\prime}$. 
This fractional Zaslavsky map is derived from fractional 
differential equation (\ref{eq7-n}) with condition (\ref{Zcond}).
For $\beta=0$ this map gives the Zaslavsky map 
(\ref{Zasl1a}) and (\ref{Zasl2a}) with $\mu =\mu^{\prime}$ and $\Omega=0$.

By analogy with Sec. 3, we can derive 
a fractional generalization of the H\'enon map 
from equation of motion (\ref{eq7-n}) with
\be \label{Ghenon-n}
G (x)= -\frac{q}{1+b} \left[ 1 + (1+b) x+ ax^2 \right] ,
\ee
where $b=-\exp(-q)$ and $T=1$.
As a result, this generalization has the form of 
Eqs. (\ref{FDR3-n}) and (\ref{FDR4-n}) with the function (\ref{Ghenon-n}).
This fractional H\'enon map is derived from the differential equations
with fractional derivative in the kicked damped term.
For $\beta=0$ this map gives the usual H\'enon map.

Computer simulations of the fractional generalizations of 
discrete maps prove that the nonlinear dynamical systems, 
which are described by the equations with fractional derivatives, 
exhibit a new type of chaotic motion. 
This type of motion can be considered 
as a fractional generalization of chaotic attractor. 
As a result, the fractional discrete maps allow to study 
a new type of attractors that are called pseudochaotic.

%%%%%%%%%%%%%%%%%%%%%%%%%%%%%%%%%%%%%%%%%%%%%%%%%%%%%%%%%%%%%%%%%%%%%%%%%%%

\section{Conclusion}

In many areas of mechanics and physics 
the problems can be reduced to the study of discrete maps. 
In particular the special case of discrete maps 
has been studied to describe properties of regular
and strange attractors. 
Under a wide range of circumstances such maps give rise to chaotic behavior. 
The suggested fractional maps can be considered as 
a fractional generalization of discrete map. 
Note that a wide class of these maps can be 
derived from kicked fractional differential equations. 
The suggested fractional Zaslavsky map and the fractional H\'enon map 
can demonstrate a chaotic behavior with a new type of attractors.
The interesting property of these fractional discrete maps is a long-term memory.
As a result, a present state evolution depends 
on all past states with the fractional power-law weights functions
$V_{\alpha}(z)$ and $W_{\alpha}(q,T,z)$ defined by equations (\ref{V}) and (\ref{Wa}). 
Note that the fractional maps are equivalent to the
correspondent fractional kicked differential equations.
An approximation for fractional derivatives of these equations is not used.
This fact can be used to study the evolution that is described 
by fractional differential equations and 
to describe pseudochaotic attractors.

%%%\input{referenc}
%%%%%%%%%%%%%%%%%%%%%%%%%%%%%%%%%%%%%%%%%%%%%%%%%%%%%%%%%%%%%%%%%%%%%%%%%%%%%%%%%

%%%%%%%%%%%%%%%%%%%%%%%%%%%%%%%%%%%%%%%%%%%%%%%%%%%%%%%%%%%%%%%%%%%%%%%%%%%%%

\end{document}